\documentclass[aps,prl,reprint,nofootinbib]{revtex4-1}
\usepackage{amsfonts,amsmath}
\usepackage{MnSymbol}
\usepackage{xspace}

\newcommand{\QKD}{QKD\xspace}
\newcommand{\Xbasis}{{\mathtt X}\xspace}
\newcommand{\Zbasis}{{\mathtt Z}\xspace}
\newcommand{\Bbasis}{{\mathtt B}\xspace}
\newcommand{\FK}{FRKL\xspace}

\newtheorem{Def}{Definition}
\newtheorem{Thrm}{Theorem}

\DeclareMathOperator{\Width}{Width}

\begin{document}
\title{Application Of McDiarmid Inequality In Finite-Key-Length Decoy-State
 Quantum Key Distribution}

\author{H. F. Chau}
\thanks{email: \texttt{hfchau@hku.hk}}
\affiliation{Department of Physics, University of Hong Kong, Pokfulam Road,
 Hong Kong}
\affiliation{Center of Theoretical and Computational Physics, University of
 Hong Kong, Pokfulam Road, Hong Kong}
\date{\today}

\begin{abstract}
 In practical decoy-state quantum key distribution, the raw key length is
 finite.
 Thus, deviation of the estimated single photon yield and single photon error
 rate from their respective true values due to finite sample size can
 seriously lower the provably secure key rate $R$.
 Current method to obtain a lower bound of $R$ follows an indirect path by
 first bounding the yields and error rates both conditioned on the type of
 decoy used.
 These bounds are then used to deduce the single photon yield and error rate,
 which in turn are used to calculate a lower bound of the key rate $R$.
 Here I show how to directly compute a lower bound of $R$ via McDiarmid
 inequality in statistics.
 This method increases the provably secure key rate of realistic quantum
 channels by at least 30\% when the raw key length is $\approx 10^5$ to
 $10^6$.
 More importantly, this is achieved by pure theoretical analysis without
 altering the experimental setup or the post-processing method.
 In a boarder context, this work introduces powerful concentration inequality
 techniques in statistics to tackle physics problem beyond straightforward
 statistical data analysis.
\end{abstract}

\maketitle

 Quantum key distribution (\QKD) enables two trusted parties Alice and Bob to
 share a provably secure secret key by preparing and measuring quantum states
 that are transmitted through a noisy channel controlled by an eavesdropper
 Eve.
 One of the major challenges to make \QKD practical is to increase the number
 of secure bits generated per second~\cite{DLQY16}.
 That is why most \QKD experiments to date use photons as the quantum
 information carriers; and these photons come from phase randomize Poissonian
 distributed sources instead of the much less efficient single photon sources.
 In addition, decoy state method is used to combat Eve's
 photon-number-splitting attack on multiple photon events emitted from the
 Poissonian sources~\cite{Wang05,LMC05}.
 From the theoretical point of view, a more convenient figure of merit is the
 key rate, namely, the number of provably secure secret bits per average
 number of photon pulses prepared by Alice.
 This is because key rate measures the intrinsic performance of a \QKD
 protocol (in other words, the software issue) without taking the frequency of
 the pulse (which is a hardware issue) into account.

 Provably secure lower bounds of key rates (I refer them as simply as key
 rates from now on) for various \QKD schemes for the realistic situation of
 finite raw key length have been reported.
 For instance, Lim \emph{et al.}~\cite{LCWXZ14} computed the key rates of a
 certain implementation of the BB84 \QKD scheme~\cite{BB84} using three types
 of decoy; recently, Chau~\cite{Chau18} extended it to the case of using more
 than three types of decoys.
 Hayashi and Nakayama studied the key rate for the BB84 scheme~\cite{HN14}.
 And Br\'{a}dler \emph{et al.} showed the key rate for a qudit-based \QKD scheme
 using up to three mutually unbiased preparation and measurement
 bases~\cite{BMFBB16}.
 Note that these key rates are found using the following three-step strategy.
 First, the yields $Q_{\Bbasis,\mu_n}$ and error rates $E_{\Bbasis,\mu_n}$
 conditioned on the preparation and measurement basis $\Bbasis$ as well as the
 photon intensity parameter $\mu_n$ used are determined by comparing the
 relevant Bob's measurement outcomes, if any, with Alice's preparation states.
 The second step is to deduce yields and error rates conditioned on the number
 of photons emitted by the source.
 Recall that for a phase randomized Poissonian photon source,
\begin{equation}
 Q_{\Bbasis,\mu_n} = \sum_{m=0}^{+\infty} \frac{\mu_n^m Y_{\Bbasis,m}
 \exp(-\mu_n)}{m!}
 \label{E:Q_mu_def}
\end{equation}
 and
\begin{equation}
 Q_{\Bbasis,\mu_n} E_{\Bbasis,\mu_n} = \sum_{m=0}^{+\infty} \frac{\mu_n^m
 Y_{\Bbasis,m} e_{\Bbasis,m} \exp(-\mu_n)}{m!} .
 \label{E:E_mu_def}
\end{equation}
 Here, $\mu_1 > \mu_2 > \cdots > \mu_k \ge 0$ are the photon intensities used
 in the decoy method with $k \ge 2$.
 Moreover, $Y_{\Bbasis,m}$ is the probability of photon detection by Bob given
 that the photon pulse sent by Alice contains $m$~photons and $e_{\Bbasis,m}$
 is the bit error rate for $m$~photon emission events prepared in the
 $\Bbasis$ basis~\cite{Wang05,LMC05,MQZL05}.
 The key rate $R$ depends on $Y_{\Bbasis,0}$, $Y_{\Bbasis,1}$ and
 $e_{\Bbasis,1}$~\cite{LCWXZ14,Wang05,LMC05,MQZL05}.
 Nevertheless, the later quantities cannot be determined precisely because
 Eqs.~\eqref{E:Q_mu_def} and~\eqref{E:E_mu_def} are under-determined systems
 of equations given $Q_{\Bbasis,\mu_n}$'s and $E_{\Bbasis,\mu_n}$'s provided
 that the number of decoys $k$ is finite.
 To make things worse, in the finite-raw-key-length (\FK) situation, the
 measured values of $Q_{\Bbasis,\mu_n}$'s and $E_{\Bbasis,\mu_n}$'s deviate
 from their true values due to finite sampling.
 Fortunately, effective lower bounds of $Y_{\Bbasis,0}$ and $Y_{\Bbasis,1}$
 as well as upper bound of $e_{\Bbasis,1}$ are
 available~\cite{LCWXZ14,Wang05,LMC05,MQZL05,Hayashi07,Chau18}.
 In the \FK situation, these bounds can be deduced with the help of Hoeffding
 inequality~\cite{Hoeffding}.
 (See, for example, Refs.~\cite{LCWXZ14,Chau18} for details.)
 The third step is to deduce $R$ from these
 bounds~\cite{LCWXZ14,Wang05,LMC05,MQZL05,BMFBB16}.

 Computing lower bound of $R$ using this indirect strategy is not satisfactory
 in the \FK situation because it is unlikely for each of the finite-size
 fluctuations in $Q_{\Bbasis,\mu_n}$'s and $E_{\Bbasis,\mu_n}$'s to decrease
 the value of the provably secure key rate.
 In fact, for a given security parameter, the worst case bounds on
 $Y_{\Bbasis,0}$ and $Y_{\Bbasis,1}$ cannot be not attained simultaneously if
 the raw key length is finite.
 (This is evident, say, from the bounds of $Y_{\Bbasis,0}$ and $Y_{\Bbasis,1}$
 given by Inequalities~(2) and~(3) in Ref.~\cite{LCWXZ14} or
 Inequalities~(12a) and~(12b) in Ref.~\cite{Chau18}.  Note that there is a
 typo in Inequality~(12b) --- the $Q_{\Bbasis,\mu_i}^{\llangle k_0-i\rrangle}$
 there should be $Q_{\Bbasis,\mu_i}^{\llangle k_0-i+1\rrangle}$.  In all
 cases, the finite-size statistical fluctuation that leads to the saturation
 of lower bound for $Y_{\Bbasis,0}$ does not cause the saturation of the lower
 bound for $Y_{\Bbasis,1}$ and vice versa.)

 It is more effective if one could directly investigate the influence of
 finite-key-length on the key rate.
 To do so, one has to go beyond the use of Hoeffding inequality to bound the
 statistical fluctuation, which only works for equally weighted sum of
 random variables that are either statistical independent or drawn from a
 finite population without replacement~\cite{Hoeffding}.
 Here I use the computation of the key rate of a specific BB84 \QKD
 protocol~\cite{BB84} that generates the raw key solely from $\Xbasis$ basis
 measurement results as an example to illustrate how to directly tackle
 statistical fluctuation in the \FK situation by means of McDiarmid
 inequality~\cite{McDiarmid} in statistics.
 The technique used here can be easily adapted to compute the key rates of
 other \QKD schemes using finite-dimensional qudits in the \FK situation.

 Recall that the error rate for this particular BB84 \QKD scheme is
 lower-bounded by~\cite{LCWXZ14,Chau18}
\begin{align}
 & p_{\Xbasis}^2 \left\{
  \vphantom{\frac{\langle Q_{\Xbasis,\mu} \rangle}{\ell_\text{raw}} \left[
   6\log_2 \frac{\chi(k)}{\epsilon_\text{sec}} + \log_2
   \frac{2}{\epsilon_\text{cor}} \right]}
  \langle \exp(-\mu) \rangle Y_{\Xbasis,0} + \langle \mu \exp(-\mu) \rangle
  Y_{\Xbasis,1} [1-H_2(e_p)] - \Lambda_\text{EC} \right. \nonumber \\
 &\quad \left. - \frac{\langle Q_{\Xbasis,\mu} \rangle}{\ell_\text{raw}}
  \left[ 6\log_2 \frac{\chi(k)}{\epsilon_\text{sec}} + \log_2
  \frac{2}{\epsilon_\text{cor}} \right] \right\} ,
 \label{E:key_rate_basic}
\end{align}
 where $p_\Xbasis$ denotes the probability that Alice (Bob) uses $\Xbasis$ as
 the preparation (measurement) basis, $\langle f(\mu)\rangle \equiv
 \sum_{n=1}^k p_{\mu_n} f(\mu_n)$ with $p_{\mu_n}$ being the probability for
 Alice to use photon intensity parameter $\mu_n$.
 Furthermore, $H_2(x) \equiv -x \log_2 x - (1-x) \log_2 (1-x)$ is the binary
 entropy function, $e_p$ is the phase error rate of the single photon events
 in the raw key, and $\Lambda_\text{EC}$ is the actual number of bits of
 information that leaks to Eve as Alice and Bob perform error correction on
 their raw bits.
 It is given by
\begin{equation}
 \Lambda_\text{EC} = \langle Q_{\Xbasis,\mu} H_2(E_{\Xbasis,\mu}) \rangle
 \label{E:information_leakage}
\end{equation}
 if they use the most efficient (classical) error correcting code to do the
 job.
 In addition, $\ell_\text{raw}$ is the raw sifted key
 length measured in bits, $\epsilon_\text{cor}$ is the upper bound of the
 chance that the final secret keys shared between Alice and Bob are different,
 Eve's information on the final key is at most
 $\epsilon_\text{sec}$~\cite{Renner05,KGR05,RGK05}, and $\chi(k)$ is a \QKD
 scheme specific factor depending on the number of photon intensities $k$
 together with the detailed security analysis used.

 For BB84, $e_p \to e_{\Zbasis,1}$ as $\ell_\text{raw} \to +\infty$.
 More importantly, the best known bound on the difference between $e_p$ and
 $e_{\Zbasis,1}$ due to finite sample size correction using properties of
 the hypergeometric distribution reported in given by~\cite{Chau18,FMC10}
\begin{align}
 e_p &\le e_{\Zbasis,1} + \bar{\gamma}(\epsilon_\text{sec} / \chi(k),
  e_{\Zbasis,1}, s_\Zbasis Y_{\Zbasis,1} \langle \mu \exp(-\mu) \rangle /
  \langle Q_{\Zbasis,\mu} \rangle, \nonumber \\
 & \qquad s_\Xbasis Y_{\Xbasis,1} \langle \mu \exp(-\mu) \rangle /
  \langle Q_{\Xbasis,\mu} \rangle)
 \label{E:e_p_bound}
\end{align}
 with probability at least $1-\epsilon_\text{sec}/\chi(k)$, where
\begin{equation}
 \bar{\gamma}(a,b,c,d) \equiv \sqrt{\frac{(c+d)(1-b)b}{c d} \ \ln \left[
  \frac{c+d}{2\pi c d (1-b)b a^2} \right]} ,
 \label{E:gamma_def}
\end{equation}
 and $s_\Bbasis$ is the number of bits that are prepared and measured in
 $\Bbasis$ basis.
 Clearly, $s_\Xbasis = \ell_\text{raw}$ and $s_\Zbasis \approx (1-
 p_\Xbasis)^2 s_\Xbasis \langle Q_{\Zbasis,\mu} \rangle / (p_\Xbasis^2
 \langle Q_{\Xbasis,\mu} \rangle)$.
 (Note that $\bar{\gamma}$ becomes complex if $a,c,d$ are too large.  This
 is because in this case no $e_p \ge e_{\Zbasis,1}$ exists with failure
 probability $a$.  I carefully picked parameters here so that $\bar{\gamma}$
 is real.)

 In the infinite-key-length limit, statistical fluctuations of
 $Q_{\Bbasis,\mu_n}$ and $E_{\Bbasis,\mu_n}$ can be ignored.
 Then based on the analysis in Ref.~\cite{Chau18} with typos corrected, one
 has
\begin{subequations}
\label{E:parameter_bounds}
\begin{align}
 Y_{\Bbasis,0} &\ge \max \left( 0, \sum_{n=1}^k a_{0n} Q_{\Bbasis,\mu_n}
  \right) \nonumber \\
 & \equiv \max \left( 0, \sum_{n=k_0}^k \frac{-Q_{\Bbasis,\mu_n} \exp[\mu_n]
  \hat{\prod}_{i\ne n} \mu_i}{\hat{\prod}_{j\ne n} [\mu_n - \mu_j]} \right) ,
 \label{E:Y0_bound}
\end{align}
\begin{align}
 Y_{\Bbasis,1} &\ge \max \left( 0, \sum_{n=1}^k a_{1n} Q_{\Bbasis,\mu_n}
  \right) \nonumber \\
 &\equiv \max \left( 0, \sum_{n=3-k_0}^k \frac{-Q_{\Bbasis,\mu_n} \exp[\mu_n]
  \hat{S}_n}{\hat{\prod}_{j\ne n} [\mu_n - \mu_j]} \right)
 \label{E:Y1_bound}
\end{align}
 and
\begin{align}
 Y_{\Zbasis,1} e_{\Zbasis,1} &\le \min \left( \frac{Y_{\Zbasis,1}}{2} ,
  \sum_{n=1}^k a_{2n} Q_{\Zbasis,\mu_n} E_{\Zbasis,\mu_n} \right) \nonumber \\
 &\equiv \min \left( \frac{Y_{\Zbasis,1}}{2} , \sum_{n=k_0}^k
  \frac{Q_{\Zbasis,\mu_n} E_{\Zbasis,\mu_n} \exp[\mu_n]
 \hat{S}_n}{\hat{\prod}_{j\ne n} [\mu_n - \mu_j]} \right) ,
 \label{E:e1_bound}
\end{align}
\end{subequations}
 where $k_0 = 1 (2)$ if $k$ is even (odd), and $\hat{\prod}_{j\ne n}$ is over
 the dummy variable $j$ from $k_0$ to $k$ but skipping $n$.
 In addition, $\hat{S}_n = \sum'' \mu_{t_1} \mu_{t_2} \cdots
 \mu_{t_{k-k_0-1}}$ where the double primed sum is over $k_0 \le t_1 < t_2 <
 \cdots < t_{k-k_0-1} \le k$ with $t_1,t_2,\dots,t_{k-k_0-1} \ne n$.
 (In other words, $a_{01} = a_{21} = 0$ if $k$ is odd and $a_{11} = 0$ if $k$
 is even.)
 Substituting Inequalities~\eqref{E:e_p_bound}
 and~\eqref{E:parameter_bounds} into Expression~\eqref{E:key_rate_basic} gives
 the following lower bound of the key rate
\begin{equation}
 \sum_{n=1}^k b_n Q_{\Xbasis,\mu_n} - p_\Xbasis^2 \left\{ \Lambda_\text{EC} +
 \frac{\langle Q_{\Xbasis,\mu} \rangle}{\ell_\text{raw}} \left[ 6\log_2
 \frac{\chi(k)}{\epsilon_\text{sec}} + \log_2 \frac{2}{\epsilon_\text{cor}}
 \right] \right\} ,
 \label{E:key_rate}
\end{equation}
 where
\begin{equation}
 b_n = p_\Xbasis^2 \left\{ \langle \exp(-\mu) \rangle a_{0n} + \langle \mu
 \exp(-\mu) \rangle a_{1n} [1-H_2(e_p)] \right\}
 \label{E:b_n_def}
\end{equation}
 provided that $Y_{\Xbasis,0}, Y_{\Xbasis,1} > 0$.
 (The cases of $Y_{\Xbasis,0}$ or $Y_{\Xbasis,1} = 0$ can be dealt with in
 the same way by changing the definition of $b_n$ accordingly.  But these
 cases are not interesting for normally they imply $R = 0$ in realistic
 channels.)
  
 Note that the worst case key rate corresponds to the situation that the spin
 flip and phase shift errors in the raw key are uncorrelated so that Alice and
 Bob cannot use the correlation information to increase the efficiency of
 entanglement distillation.
 Thus, I may separately consider statistical fluctuations in
 $Q_{\Xbasis,\mu_n}$'s, $e_p$ in the \FK situation.
 This can be done by using McDiarmid inequality.
 Actually, this inequality was first proven using martingale technique in
 Ref.~\cite{McDiarmid} for the case of statistically independent random
 variables.
 The version I use here is the extension to statistically dependent random
 variables reported in Ref.~\cite{McDiarmid2}.  (See also a closely related
 version in Ref.~\cite{McDiarmid1}.)

\begin{Thrm}[McDiarmid]
 \label{Thrm:McDiarmid}
 Let ${\mathbf W} = (W_1,\ldots,W_n)$ be a family of possibly statistically
 dependent random variables with $W_i$ taking values in the set
 ${\mathcal W}_i$ for all $i$.
 Let $f$ be a bounded real-valued function of ${\mathbf W}$.
 For a fixed $i=1,2,\ldots, n$, let $w_i\in {\mathcal W}_i$ and set
 \begin{align}
  & \hat{r}^2_i(w_1,\ldots,w_{i-1}) \nonumber \\
  ={} & \sup \{ | E[f({\mathbf W}) \mid W_i = w_i, B_i] - \nonumber \\
  & \quad E[f({\mathbf W}) \mid W_i = w'_i, B_i] |^2 \colon w_i, w'_i \in
  {\mathcal W}_i \} ,
  \label{E:rangesquared_def}
 \end{align}
 where $E[f]$ is the expectation value of $f$, and $B_i$ denotes the
 conditions $W_k = w_k$ for $k=1,\ldots,i-1$.
 Further set $\hat{r}^2 = \sup \sum_{i=1}^n \hat{r}_i^2$, where the supremum
 is over all ${\mathbf w} \in \prod {\mathcal W}_i$.
 Then
 \begin{subequations}
 \label{E:McDiarmid}
 \begin{equation}
  \Pr(f({\mathbf W}) - E[f({\mathbf W})] \ge \delta) \le \exp (-2\delta^2 /
  \hat{r}^2)
  \label{E:McDiarmid1}
 \end{equation}
 and
 \begin{equation}
  \Pr(f({\mathbf W}) - E[f({\mathbf W})] \le -\delta) \le \exp (-2\delta^2 /
  \hat{r}^2)
  \label{E:McDiarmid2}
 \end{equation}
 \end{subequations}
 for any $\delta > 0$.
\end{Thrm}

 From the R.H.S. of Inequalities~\eqref{E:e_p_bound}
 and~\eqref{E:parameter_bounds}, I obtain $e_{\Zbasis,1} \le (\sum_{n=1}^k
 a_{2n} Q_{\Zbasis,\mu_n} E_{\Zbasis,\mu_n}) / (\sum_{n=1}^k a_{1n}
 Q_{\Zbasis,\mu_n})$.
 A naive way to study the statistical fluctuation of $e_{\Zbasis,1}$ is to
 regard $Q_{\Zbasis,\mu_n}$'s and $Q_{\Zbasis,\mu_n} E_{\Zbasis,\mu_n}$'s as
 random variables and directly apply Theorem~\ref{Thrm:McDiarmid} to the
 R.H.S. of the above inequality.
 However, it does not work for the R.H.S. of this inequality need not be
 bounded.
 Instead, I first write $Q_{\Zbasis,\mu_n} = \sum_j \tilde{W}_{nj} /
 \tilde{s}_{\Zbasis,\mu_n}$ where $\tilde{s}_{\Zbasis,\mu_n}$ is the number of
 photon pulses that Alice prepares using photon intensity $\mu_n$ and that
 Alice prepares and Bob tries to measure (but may or may not have detection)
 in $\Zbasis$ basis.
 In addition, $\tilde{W}_{nj}$ denotes the possibly correlated random variable
 whose value is $1$ ($0$) if the $j$th photon pulse among the
 $\tilde{s}_{\Zbasis,\mu_n}$ photon pulses is (not) detected by Bob.
 Clearly, $\tilde{s}_{\Zbasis,\mu_n} \approx T p_\Zbasis^2 p_{\mu_n}$ with $T$
 being the total number of photon pulses sent by Alice and $p_\Zbasis = 1 -
 p_\Xbasis$ is the probability for Alice (Bob) to prepare (measure) in the
 $\Zbasis$ basis.
 Since $s_\Zbasis \approx T p_\Zbasis^2 \langle Q_{\Zbasis,\mu} \rangle$, I
 arrive at
\begin{align}
 Y_{\Zbasis,1} \ge \sum_{n=1}^k a_{1n} Q_{\Zbasis,\mu_n} &= \frac{\langle
  Q_{\Zbasis,\mu} \rangle}{s_\Zbasis} \sum_{n=1}^k \left[
  \frac{a_{1n}}{p_{\mu_n}} \left( \sum_j \tilde{W}_{nj} \right) \right]
  \nonumber \\
 &= \frac{\langle Q_{\Zbasis,\mu} \rangle}{s_\Zbasis} \sum_{i=1}^{s_\Zbasis}
  W_{\Zbasis,i} .
 \label{E:Q_Zbasis_random_variables_expression}
\end{align}
 Here $W_{\Zbasis,i}$ is the random variable that takes the value $a_{1n} /
 p_{\mu_n}$ if the $i$th photon pulse that are prepared by Alice and then
 successfully measured by Bob both in the $\Zbasis$ basis is in fact prepared
 using photon intensity $\mu_n$.
 Recall that Eve knows the number of photons in each pulse and may act
 accordingly.
 However, she does not know the photon intensity parameter used in each pulse
 and the preparation basis until the pulse is measured by Bob.
 Hence, $W_{\Zbasis,n}$'s may be correlated.
 Actually, the most general situation is that $W_{\Zbasis,n}$'s are drawn from
 a larger population without replacement.
 That is to say, these random variables obey the multivariate hypergeometric
 distribution.

 For multivaritate hypergeometric distribution, $\hat{r}^2$ in
 Eq.~\eqref{E:McDiarmid} of Theorem~\ref{Thrm:McDiarmid} is very difficult to
 compute.
 Fortunately, it can be upper-bounded as follows.
 Inspired by Ref.~\cite{McDiarmid1}, I define the following.
\begin{Def}
 Let ${\mathbf W} = (W_1,\dots,W_n)$ be a sequence of random variables.
 Denote $p_i(W_i)$ the marginal probability distribution of each $W_i$.
 The sequence is said to be
 \emph{centering with respect to a real-valued function} $f({\mathbf W})$ if
 $\hat{r}_i^2(w_1,\dots,w_{i-1})$ is upper-bounded by the R.H.S. of
 Eq.~\eqref{E:rangesquared_def} when all $W_i$'s are statistically independent
 and follow the probability distribution $p_i(W_i)$.
 \label{Def:centering_property}
\end{Def}

 It is straightforward to check that multivariate hypergeometrically
 distributed $W_{\Zbasis,i}$'s form a centering sequence with respect to the
 function $\sum_{i=1}^{s_\Zbasis} W_{\Zbasis,i}$.
 As a consequence, Theorem~\ref{Thrm:McDiarmid} implies that the true value of
 $\sum_{n=1}^k a_{1n} Q_{\Zbasis,\mu_n}$ is less than the observed value by
 $\langle Q_{\Zbasis,\mu} \rangle \left[ \ln (1/\epsilon_\Zbasis) / 2s_\Zbasis
 \right]^{1/2} \Width(\{a_{1n} / p_{\mu_n} \}_{n=1}^k)$ with probability at
 most $\epsilon_\Zbasis$, where $\Width(S)$ of a bounded set $S$ of real
 numbers is defined as $\sup S - \inf S$.

 By the same token, $Y_{\Zbasis,1} e_{\Zbasis,1} \le \sum_{n=1}^k a_{2n}
 Q_{\Zbasis,\mu_n} E_{\Zbasis,\mu_n} = \langle Q_{\Zbasis,\mu} E_{\Zbasis,\mu}
 \rangle \sum_{i=1}^{s^\text{e}_\Zbasis} W_{\Zbasis,i}^\text{e} /
 s^\text{e}_\Zbasis$, where $W^\text{e}_{\Zbasis,i}$ is a random variable
 taking value of $a_{2n} / p_{\mu_n}$ if the $i$th photon pulse that is
 prepared and successfully measured in the $\Zbasis$ basis and that the
 measurement result is different from the preparation (in which there are
 totally $s^\text{e}_\Zbasis \approx T p_\Zbasis^2 \langle Q_{\Zbasis,\mu}
 E_{\Zbasis,\mu} \rangle$ such pulses) is in fact prepared using photon
 intensity $\mu_n$.
 Hence, with probability at most $\epsilon_\Zbasis^\text{e}$, the true value
 of $\sum_{n=1}^k a_{2n} Q_{\Zbasis,\mu_n} E_{\Zbasis,\mu_n}$ is greater than
 the observed value by $\langle Q_{\Zbasis,\mu} E_{\Zbasis,\mu} \rangle \left[
 \ln (1/\epsilon^\text{e}_\Zbasis) / 2s^\text{e}_\Zbasis \right]^{1/2}
 \Width(\{ a_{2n} / p_{\mu_n} \}_{n=1}^k) = \left[ \langle Q_{\Zbasis,\mu}
 \rangle \langle Q_{\Zbasis,\mu} E_{\Zbasis,\mu} \rangle \ln
 (1/\epsilon^\text{e}_\Zbasis) / 2s_\Zbasis \right]^{1/2} \Width (\{ a_{2n} /
 p_{\mu_n} \}_{n=1}^k)$.

 Since $W_{\Zbasis_i}^\text{e}$ and $W_{\Zbasis,j}$ are positively correlated,
 with probability at least $1 - \epsilon_\Zbasis - \epsilon_\Zbasis^\text{e} -
 \epsilon_{\bar\gamma}$, the phase error rate $e_p$ is upper-bounded by the
 R.H.S. of Inequality~\eqref{E:e_p_bound} where
\begin{widetext}
\begin{equation}
 e_{\Zbasis,1} = \frac{\sum_{n=1}^k a_{2n} Q_{\Zbasis,\mu_n}
 E_{\Zbasis,\mu_n} + \left[ \langle Q_{\Zbasis,\mu} \rangle \langle
 Q_{\Zbasis,\mu} E_{\Zbasis,\mu} \rangle \ln (1/\epsilon^\text{e}_\Zbasis) / 2
 s_\Zbasis \right]^{1/2} \Width (\{ a_{2n} / p_{\mu_n}
 \}_{n=1}^k)}{\sum_{n=1}^k a_{1n} Q_{\Zbasis,\mu_n} - \langle Q_{\Zbasis,\mu}
 \rangle \left[ \ln (1/\epsilon_\Zbasis) / 2 s_\Zbasis \right]^{1/2}
 \Width(\{a_{1n} / p_{\mu_n} \}_{n=1}^k)} .
 \label{E:e_Z1_bound}
\end{equation}
\end{widetext}

 To study the statistical fluctuation of $R$, it remains to consider the
 fluctuation of $Q_{\Xbasis,\mu_n}$ in the first term in
 Expression~\eqref{E:key_rate}.
 (Although the second term also depends on $Q_{\Xbasis,\mu_n}$'s implicitly
 through $\Lambda_\text{EC}$, statistical fluctuation is absent from this
 term.
 This is because $\Lambda_\text{EC}$ is the amount of information leaking to
 Eve during classical post-processing of the measured raw bits.
 Thus, it depends on the observed values of $Q_{\Xbasis,\mu_n}$'s and
 $E_{\Xbasis,\mu_n}$'s instead of their true values.)
 Using the same technique as in the estimation of statistical fluctuation in
 $e_p$, the first term of Expression~\eqref{E:key_rate} can be rewritten as
 $\langle Q_{\Xbasis,\mu} \rangle \sum_{i=1}^{s_\Xbasis} W_{\Xbasis,i}$ where
 $W_{\Xbasis,i}$'s are multivariate hypergeometrically distributed random
 variables each taken values in the set $\{ b_n / p_{\mu_n} \}_{n=1}^k$.
 Here $b_n$ is given by Eq.~\eqref{E:b_n_def} with $e_p$ equals the R.H.S. of
 Inequality~\eqref{E:e_p_bound} where $e_{\Zbasis,1}$ satisfies
 Eq.~\eqref{E:e_Z1_bound}.
 Theorem~\ref{Thrm:McDiarmid} implies that due to statistical fluctuation, the
 true value of the first term in Expression~\eqref{E:key_rate} is lower than
 the observed value by $\langle Q_{\Xbasis,\mu} \rangle \left[ \ln
 (1/\epsilon_\Xbasis) / (2 s_\Xbasis) \right]^{1/2} \Width (\{ b_n / p_{\mu_n}
 \}_{n=1}^k)$ with probability at most $\epsilon_\Xbasis$.

 Putting everything together and by setting $\epsilon_\Xbasis =
 \epsilon_\Zbasis = \epsilon_\Zbasis^\text{e} = \epsilon_{\bar{\gamma}} =
 \epsilon_\text{sec} / \chi(k)$, I conclude that the secret key rate $R$
 satisfies
\begin{widetext}
\begin{equation}
 R = \sum_{n=1}^k b_n Q_{\Xbasis,\mu_n} - \langle Q_{\Xbasis,\mu} \rangle
 \left\{ \frac{\ln[\chi(k) / \epsilon_\text{sec}]}{2s_\Xbasis} \right\}^{1/2}
 \Width(\{ \frac{b_n}{p_{\mu_n}} \}_{n=1}^k) - p_\Xbasis^2 \left\{ \langle
 Q_{\Xbasis,\mu} H_2(E_{\Xbasis,\mu}) \rangle + \frac{\langle Q_{\Xbasis,\mu}
 \rangle}{s_\Xbasis} \left[ 6\log_2 \frac{\chi(k)}{\epsilon_\text{sec}} +
 \log_2 \frac{2}{\epsilon_\text{cor}} \right] \right\} ,
 \label{E:finite-size_key_rate}
\end{equation}
\end{widetext}
 where $b_n = b_n(e_p)$ is given by Eq.~\eqref{E:b_n_def}.
 Here $e_p$ equals the R.H.S. of Inequality~\eqref{E:e_p_bound} with
 $e_{\Zbasis,1}$ given by Eq.~\eqref{E:e_Z1_bound}.
 Interestingly, $\chi(k) = 9 = 4¡Ï1+4$ is independent on the number of photon
 intensities $k$ used.
 (Here the first number $4$ comes from the generalized chain rule for smooth
 entropy in Ref.~\cite{LCWXZ14}, the number $1$ comes from the finite-size
 correction of the raw key in Eq.~(B1) of Ref.~\cite{LCWXZ14}, and the last
 number $4$ comes from $\epsilon_\Xbasis, \epsilon_\Zbasis,
 \epsilon_\Zbasis^\text{e}$ and $\epsilon_{\bar{\gamma}}$ through the use of
 McDiarmid inequality~\cite{McDiarmid2} and hypergeometric distribution bound
 in Ref.~\cite{FMC10}.)
 Although $\chi$ does not depend on $k$ for this method, it does not mean that
 one could use arbitrarily large number of photon intensities as decoys
 without adversely affecting the key rate for a fixed finite $s_\Xbasis$.
 The reason is that $\Width (\{ a_{1n} / p_{\mu_n} \}_{n=1}^k \})$, $\Width
 (\{ a_{2n} / p_{\mu_n} \}_{n=1}^k \})$ and $\Width (\{ b_n / p_{\mu_n}
 \}_{n=1}^k \})$ diverge as $k\to +\infty$ due to divergence of $a_{1n}$,
 $a_{2n}$ and $b_n$~\cite{Chau18} as well as the decrease in $\min \{
 p_{\mu_n} \}_{n=1}^k$.
 Recall that computing $a_{1n}$, $a_{2n}$ and $b_n$ is numerically stable and
 with minimal lost in precision if $\mu_n - \mu_{n+1} \gtrsim 0.1$ for $n=1,2,
 \dots,k-1$~\cite{Chau18}.
 This means the number of photon intensities $k$ used in practice should be
 $\lesssim 10$.
 
 To evaluate the performance of this new key rate formula in realistic
 situation, I consider the quantum channel with $Q_{\Bbasis,\mu} \approx
 (1+p_\text{ap})(2p_\text{dc}+\eta_\text{sys}\mu)$ and $Q_{\Bbasis,\mu}
 E_{\Bbasis,\mu} \approx (1+p_\text{ap}) p_\text{dc} + (e_\text{mis}
 \eta_\text{ch} + p_\text{ap} \eta_\text{sys} / 2)\mu$ for $0\le \mu\le 1$,
 which is a commonly used channel model for dedicated-optical-fiber-based
 \QKD experiments.
 Here I fix after pulse probability $p_\text{ap} = 4\times 10^{-2}$, dark
 count probability $p_\text{dc} = 6\times 10^{-7}$, error rate of the optical
 system $e_\text{mis} = 5\times 10^{-3}$, transmittances of the fiber and the
 system $\eta_\text{ch} = 1 \times 10^{-2}$ and $\eta_\text{sys} = 1\times
 10^{-3}$.
 These parameters are obtained from optical fiber experiment on a 100~km long
 fiber in Ref.~\cite{WLGHZG12}; and have been used in
 Refs.~\cite{LCWXZ14,Chau18} to study the performance of decoy-state \QKD in
 the \FK situation.
 I also follow Refs.~\cite{LCWXZ14,Chau18} by using the following security
 parameters:
 $\epsilon_\text{cor} = \kappa = 10^{-15}$, where $\epsilon_\text{sec} =
 \kappa \ell_\text{final}$ with $\ell_\text{final} \approx R s_\Xbasis /
 (p_\Xbasis^2 \langle Q_{\Xbasis,\mu} \rangle)$ is the length of the final
 key measured in bits.
 Note that $\kappa$ can be interpreted as the secrecy leakage per final secret
 bit.

\begin{table}[th]
 \centering
 \begin{tabular}{||c|c|c|c|c|c|c|c|c||}
  \hline\hline
   & \multicolumn{2}{c|}{$k=3$} & \multicolumn{2}{c|}{$k=4$} &
   \multicolumn{2}{c|}{$k=5$} & \multicolumn{2}{c||}{$k=6$} \\
  \cline{2-9}
  $s_\Xbasis$ & $R'_{-5}$ & $R_{-5}$ & $R'_{-5}$ & $R_{-5}$ & $R'_{-5}$ &
   $R_{-5}$ & $R'_{-5}$ & $R_{-5}$ \\
  \hline
  $10^5$ & $0.052$ & $0.102$ & $0.027$ & $0.070$ & $0.000$ & $0.027$ &
   $0.000$ & $0.004$ \\
  \hline
  $10^6$ & $0.294$ & $0.388$ & $0.194$ & $0.327$ & $0.100$ & $0.212$ &
   $0.055$ & $0.129$ \\
  \hline
  $10^7$ & $0.687$ & $0.779$ & $0.573$ & $0.756$ & $0.421$ & $0.596$ &
   $0.259$ & $0.410$ \\
  \hline
  $10^8$ & $1.11$ & $1.17$ & $1.04$ & $1.21$ & $0.929$ & $1.11$ & $0.624$ &
   $0.874$ \\
  \hline
  $10^9$ & $1.51$ & $1.53$ & $1.57$ & $1.64$ & $1.46$ & $1.63$ & $1.08$ &
   $1.33$ \\
  \hline
  $10^{10}$ & $1.87$ & $1.88$ & $1.97$ & $2.06$ & $1.94$ & $2.12$ & $1.72$ &
   $1.78$ \\
  \hline
  $10^{11}$ & $2.20$ & $2.21$ & $2.32$ & $2.37$ & $2.46$ & $2.54$ & $2.18$ &
   $2.22$ \\
  \hline\hline
 \end{tabular}
 \caption{Comparison between the state-of-the-art key rate $R' \equiv R'_{-5}
  \times 10^{-5}$ in Ref.~\cite{Chau18} with the key rate in
  Eq.~\eqref{E:finite-size_key_rate} (or more precisely $R_{-5} \equiv \max
  (0,R \times 10^{-5})$) for the dedicated quantum channel used in
  Refs.~\cite{LCWXZ14,Chau18}.
  These rate are optimized using the method stated in the main text.
 \label{T:keyrates}}
\end{table}

 Table~\ref{T:keyrates} compares the optimized key rates for the
 state-of-the-art method reported recently Eq.~(3) of Ref.~\cite{Chau18} with
 Eq.~\eqref{E:finite-size_key_rate} for various $s_\Xbasis$ and $k$.
 The optimized rates are found by fixing the minimum photon intensity to
 $1\times 10^{-6}$, while maximizing over $p_\Xbasis$ as well as all other
 photon intensities $\mu_n$'s and all the $p_{\mu_n}$'s.
 The table clearly shows that using McDiarmid inequality improves the
 optimized key rates in all cases.
 In terms of the percentage increase in key rate, the smaller the raw key
 length $s_\Xbasis$, the better the improvement.
 (And the improvement vanishes as $s_\Xbasis\to +\infty$.)
 For $s_\Xbasis \approx 10^5 - 10^6$, the improvement is at least $30\%$.
 This improvement is of great value in practical \QKD because the
 computational and time costs for classical post-processing can be quite high
 when the raw key length $s_\Xbasis$ is long.
 More importantly, the McDiarmid inequality method reported here is effective
 to increase the key rate of real or close to real time on demand generation
 of the secret key --- an application that is possible in near future with the
 advancement of laser technology.

 In addition to \QKD, powerful concentration inequalities in statistics such
 as McDiarmid inequality could also be used beyond straightforward statistical
 data analysis.
 One possibility is to use it to construct model independent test for physics
 experiments that involve a large number of parameters but with relatively
 few data points.

\begin{acknowledgments}
 This work is supported by the RGC grant~17304716 of the Hong Kong SAR
 Government.
 I would like to thank Joseph K.~C. Ng for his discussions on the McDiarmid
 inequality and K.-B. Luk for his discussion on potential applications of
 McDiarmid inequality in physics.
\end{acknowledgments}

\bibliographystyle{apsrev4-1}

\bibliography{qc75.1}

\end{document}